\documentclass[twocolumn,showpacs,preprintnumbers,amsmath,amssymb]{revtex4}

\usepackage{graphicx}
\usepackage{dcolumn}
\usepackage{bm}

\begin{document}


\title{Thermal Conductivity of the Pyrochlore Superconductor KOs$_2$O$_6$: \\
Strong Electron Correlations and Fully Gapped Superconductivity}

\author{Y.~Kasahara,$^1$ Y.~Shimono,$^1$ T.~Shibauchi,$^1$ 
Y.~Matsuda,$^{1,2}$ S.~Yonezawa,$^2$ Y.~Muraoka,$^2$ and Z.~Hiroi$^2$}
\affiliation{
$^1$Department of Physics, Kyoto University, Sakyo-ku, 
Kyoto 606-8502, Japan \\
$^2$Institute for Solid State Physics, University of Tokyo, 
Kashiwa, Chiba 277-8581, Japan
}%

\date{\today}

\begin{abstract}
To elucidate the nature of the superconducting ground state of the 
geometrically frustrated pyrochlore KOs$_2$O$_6$ ($T_c$~=~9.6~K), 
the thermal conductivity was measured down to low temperatures 
($\sim T_c/100$). 
We found that the quasiparticle mean free path is strikingly enhanced 
below a transition at $T_p=7.8$~K, indicating enormous electron 
inelastic scattering in the normal state. 
In magnetic fields, the conduction at $T \rightarrow $~0~K is nearly 
constant up to $\sim0.4H_{c2}$, in contrast with the rapid growth expected 
for superconductors with an anisotropic gap. 
This unambiguously indicates a fully gapped superconductivity, 
in contrast to the previous studies. 
These results highlight that KOs$_2$O$_6$ is unique among 
superconductors with strong electron correlations.

\end{abstract}


\maketitle 

Geometrically frustrated systems have recently become the subject of 
intense theoretical and experimental study. 
A fundamental question for such systems is the nature of the ground state. 
For itinerant electron systems, it has been argued that geometrically 
frustrated lattice gives rise to exotic phenomena including heavy fermion 
states \cite{kondo}, metal-insulator transitions \cite{fujimoto}, 
and the anomalous Hall effect \cite{taguchi}. 
The pyrochlore system is an ideal system for such studies, since the network 
of the relevant metal atoms consists of corner sharing tetrahedra. 
Very recently, superconductivity has been discovered in the 
$\beta$-pyrochlore oxides $R$Os$_2$O$_6$ ($R$~=~Cs, Rb, and K) 
\cite{yonezawa1,yonezawa2,hiroi1,yonezawa3,kazanov}. 
These compounds have attracted great interest because 
the geometrical frustration inherent to the crystal structure 
may give rise to an exotic superconducting state. 

CsOs$_2$O$_6$ ($T_c=3.3$~K) and RbOs$_2$O$_6$ ($T_c=6.3$~K) 
show rather usual behavior. In both compounds, $T^2$-dependent 
resistivity $\rho$ is observed and the upper critical fields $H_{c2}$ 
are below the Pauli limit. 
A jump $\Delta C$ in the specific heat at $T_c$ implies 
superconductivity in the intermediate coupling regime. 
Several experiments indicate the conventional BCS $s$-wave 
superconductivity 
\cite{koda,khasanov,bruhwiler1,bruhwiler2,magishi}. 

In sharp contrast, the behavior of KOs$_2$O$_6$ with the highest $T_c$ 
(=~9.6~K) appears to be highly unusual. 
KOs$_2$O$_6$ exhibits an extremely low-energy large rattling motion 
of the K ions 
in an oversized cage forming a three dimensional skeleton \cite{pickett}. 
In the normal state, the resistivity exhibits a strong convex $T$-dependence 
from just above $T_c$ extending to room temperature, 
indicating anomalous electron scattering \cite{hiroi1}. 
Specific heat measurement has revealed an unusually large mass enhancement 
with a Sommerfeld coefficient of $\gamma=$70-110~mJ/K$^2$mol, 
which is strongly enhanced from the band calculation value of 
9.8~mJ/K$^2$mol \cite{hiroi2,bruhwiler3}. 
In addition, strong coupling superconductivity has been suggested 
based on the large $\Delta C$ \cite{hiroi2,bruhwiler2}. 
With decreasing $T$, $H_{c2}$ increases linearly even below 1~K, 
showing no saturation \cite{ohmichi}. 
Moreover, $H_{c2}\simeq$~32~T at $T \rightarrow 0$, exceeding 
the apparent Pauli limited value $\sim 18$~T. 
Measurements of $\lambda$ by $\mu$SR \cite{koda} 
and NMR relaxation rate $T_1^{-1}$ \cite{arai} 
suggest a very anisotropic gap function, 
although low-temperature measurements below 2~K are lacking. 
Very recently, a first-order phase transition, which occurs at 
$T_p \sim 7.5$~K below $T_c$, has been reported \cite{hiroi2}. 
This transition is insensitive to magnetic field and it is suggested 
that it can be associated with the rattling of the K atoms, 
though the details are unknown.  

Thus a major outstanding question is how geometrical frustration and 
the rattling motion affect the superconducting state in KOs$_2$O$_6$, 
including the microscopic pairing mechanism. 
To clarify this issue, a detailed study of the low-energy 
quasiparticle (QP) properties is of primary importance. 
We here report the QP and phonon transport 
probed by the thermal conductivity. 
We observed anomalous QP dynamics in the superconducting state 
suggesting inherent strong electron correlations 
that usually prefer anisotropic superconductivity. 
However, we found strong evidence for an isotropic gap.  
These contrasting results highlight the distinct superconducting state 
in KOs$_2$O$_6$. 

Thermal conductivity $\kappa$ was measured by a standard steady-state method 
in single crystals of KOs$_2$O$_6$ with cubic structure 
grown by the technique described in \cite{hiroi2}. 
The contact resistance was less than 0.1~$\Omega$ at low temperatures. 
The magnetic field $H$ was applied parallel to the current direction to 
minimize the vortex motion. 
As shown in the inset of Fig.~1, 
anomalous convex resistivity $\rho(T)$ from just above $T_c$ up to 300~K 
is observed \cite{hiroi1}. 
The main panel of Fig.~1 depicts the $T$-dependence of $\kappa$, 
together with $C/T$ in zero field. 
Both $\kappa$ and $C/T$ exhibit distinct anomalies at $T_c$ and $T_p$. 
Upon entering the superconducting state, $\kappa(T)$ changes its slope 
with a cusp at $T_c$ and decreases with decreasing $T$, followed by a  
kink near $T_p$.
Below $T_p$, $\kappa(T)$ increases gradually, peaks at $\sim7$~K, 
and then decreases rapidly down to $\sim$3~K. 
Below $\sim$~3~K, it decreases gradually. 
The Lorentz number $L=\kappa\rho/T=1.1L_0$ at $T_c$ is close to 
the Sommerfeld value $L_0=2.44\times10^{-8} ~\Omega$W/K$^2$, 
indicating substantial electronic contribution in 
the heat conduction near $T_c$. 

\begin{figure}[t]	
\includegraphics[width=80mm]{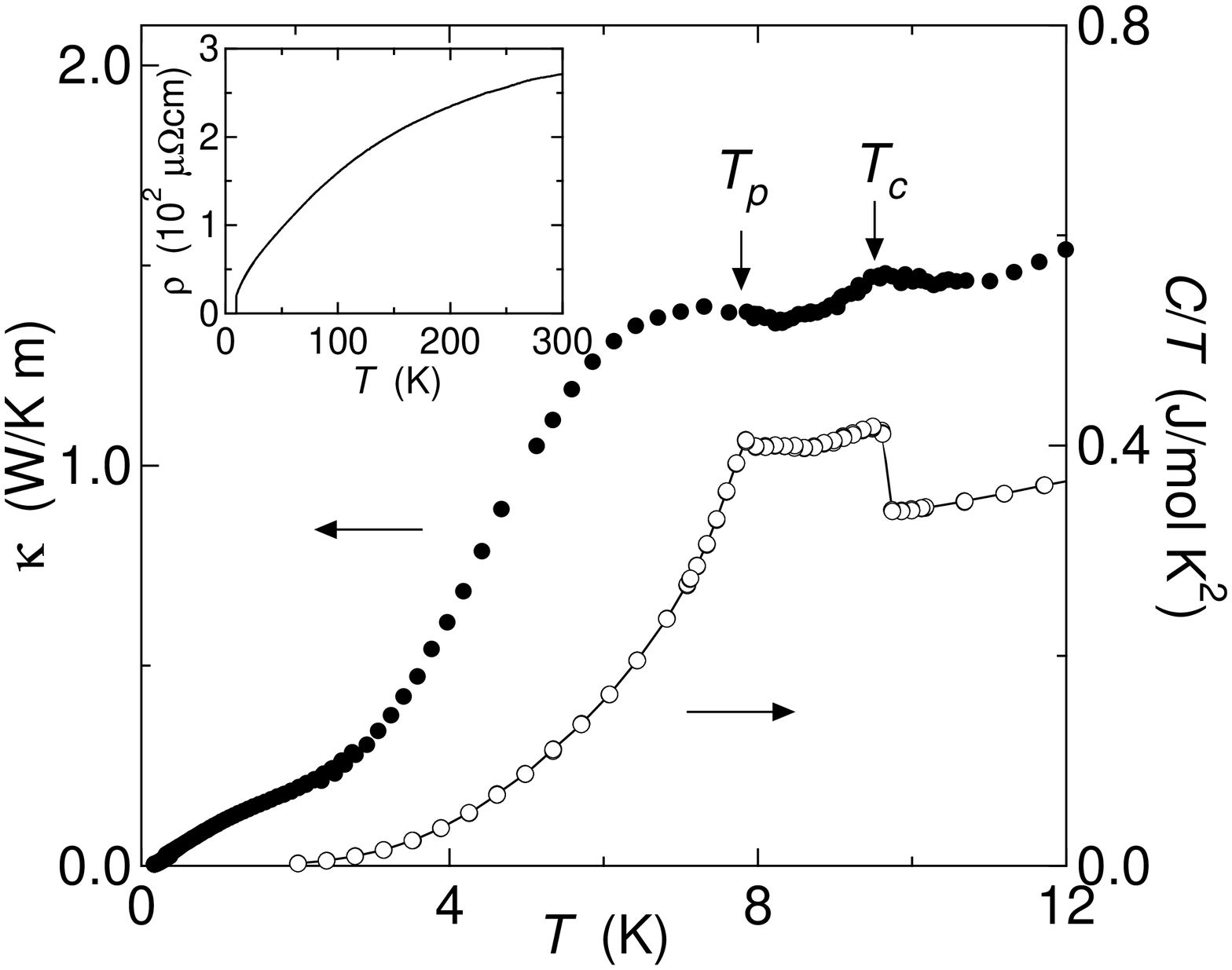}
\caption{
Zero-field thermal conductivity and specific heat
in a KOs$_2$O$_6$ crystal ($0.3 \times 0.37 \times 0.88 $~cm$^3$). 
Inset: Temperature dependence of the resistivity. }
\end{figure}

In Fig.~2(a), the $T$-dependence of $\kappa(T)/T$ for several magnetic fields 
is plotted. In zero field, $\kappa/T$ displays a steep increase below $T_p$ 
and exhibits a pronounced maximum at $\sim$~6~K. 
At very low temperatures, $\kappa(T)/T$ decreases rapidly with 
decreasing $T$ after showing a second maximum at $\sim$~0.8~K. 
The peak structure in $\kappa/T$ below $T_c(H)$ disappears above 1~T, 
indicating that the enhancement is readily suppressed by a magnetic field. 

\begin{figure}[t]
\includegraphics[width=85mm]{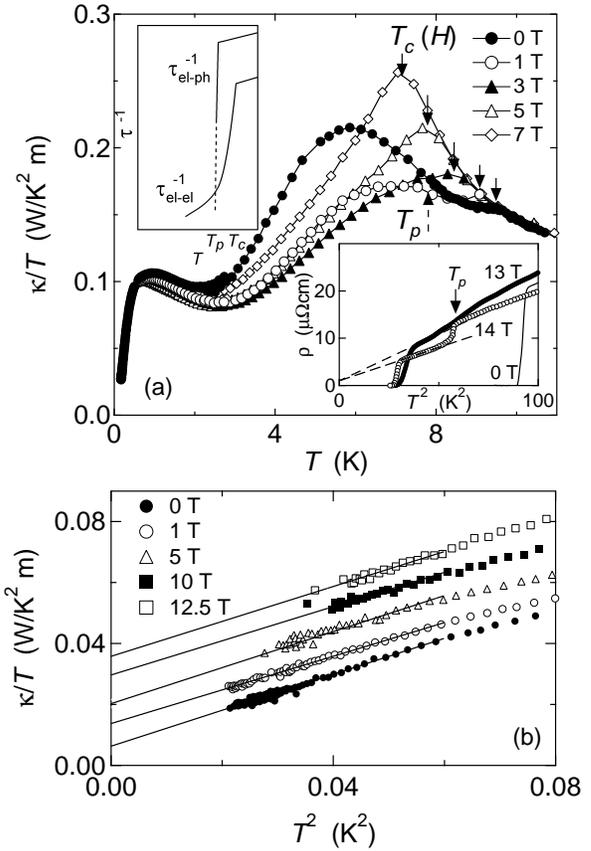}
\caption{
(a) $T$-dependence of $\kappa/T$ in various magnetic fields. 
The superconducting transition temperatures $T_c(H)$ 
are shown by arrows. $T_p$ (dashed arrow) is independent of $H$. 
Upper inset: schematic figure of $T$-dependence of $\tau_{\rm el-ph}^{-1}$ 
and $\tau_{\rm el-el}^{-1}$. 
Lower inset: $\rho$ vs $T^2$ at 0 and 13~T (solid circles). 
Data for a different crystal at 14~T are also plotted (open circles). 
The dashed lines represent $\rho(T)=\rho_0+AT^2$. 
(b) $\kappa/T$~vs~$T^2$ at low temperatures. 
The solid lines are linear fits. 
}
\end{figure}

We first discuss the origin of the double peak structure in 
$\kappa/T$ below $T_p$ in zero field. 
The maximum in $\kappa/T$ at $\sim 0.8$~K is attributed to the phonon peak. 
A peak appears when the phonon conduction grows rapidly 
at very low temperatures and is limited by the sample boundaries. 
Such a phonon peak well below $T_c$ has been reported for Nb and 
LuNi$_2$B$_2$C \cite{kes,boakinin2}. 
Below $\sim 250$~mK, the phonon contribution $\kappa_{ph}=b T^3$ can be 
separated from the electronic one $\kappa_{e}=a T$ by a fit
$\kappa/T=a+bT^2$ in Fig.~2(b). 
From the coefficient $b$, we obtain a reasonable value 
of the acoustic phonon velocity $v_s \simeq$ 4700~m/s \cite{boakinin2}.

The other peak in $\kappa/T$ at $\sim $~6~K is most likely to be of 
electronic origin, because (1) the appearance of phonon peaks twice 
below $T_c$ is highly unlikely, and (2) the electronic contribution is 
substantial near $T_c$. 
The electronic heat conduction is described by 
$\kappa_e/T\sim N(0)v_F^2\tau_e$, where $N(0)$, $v_F$, and $\tau_e$ are 
the QP density of states (DOS) at the Fermi level, Fermi velocity, 
and QP scattering time, respectively. Therefore the enhancement 
of $\kappa_e/T$ is a strong indication that 
a striking enhancement of $\tau_e$ occurs overcoming the reduction of
$N(0)$ in the superconducting state of KOs$_2$O$_6$. 
Recent microwave measurements also show the enhancement of 
the QP conductivity \cite{watanabe} consistent with our observation. 

We here examine the $T$-dependence of $\tau_e$ quantitatively below $T_c$, 
assuming $\kappa \simeq \kappa_e$. 
In the temperature range $T_p<T<T_c$, $\kappa/T$ decreases from 
that extrapolated above $T_c$, indicating that 
the thermal conductivity is mainly 
determined by the change of $N(0)$. 
We note that $N(0)$ is little affected by the phase 
transition at $T_p$, 
since $H_{c2}$ changes only slightly and 
no discernible change of the penetration depth is observed 
\cite{koda,hiroi2}. 
Therefore the enhancement of $\tau_e$ below $T_p$ is responsible for 
the increase of $\kappa/T$. 
While $\kappa/T(0.6T_c)$ is enhanced nearly 1.5 times compared with 
$\kappa/T(T_c)$, $N(0)$ at $T=0.6~T_c$ is reduced to $\sim$~1/4 
of $N(0)$ at $T_c$ according to $\mu$SR measurements \cite{koda}. 
Hence $\tau_e$ is enhanced by 6 times at 0.6~$T_c$, indicating a 
remarkable enhancement in the superconducting state. 

A rapid increase of $\tau_e$ indicates the presence of 
strong inelastic scattering originating from an interaction that 
has developed a gap \cite{yu}. 
We then expect two sources of inelastic scattering rate in KOs$_2$O$_6$, 
electron-phonon $\tau_{\rm el{\mathchar"712D}ph}^{-1}$ 
and electron-electron scattering rate 
$\tau_{\rm el{\mathchar"712D}el}^{-1}$, so that $\tau_e$ 
can be expressed as 
$\tau_e^{-1}=\tau_{\rm imp}^{-1}+\tau_{\rm el{\mathchar"712D}el}^{-1}
+\tau_{\rm el{\mathchar"712D}ph}^{-1}$,
where $\tau_{\rm imp}^{-1}$ is the impurity scattering rate. 
From the convex $\rho(T)$ and the large rattling motion of the K ions, 
we speculate that above $T_p$, the electron-phonon scattering dominates, 
$\tau_{\rm el{\mathchar"712D}ph}^{-1}\gg\tau_{\rm imp}^{-1}, 
\tau_{\rm el{\mathchar"712D}el}^{-1}$. 
Since the enhancement of phonon conductivity occurs at very low 
temperatures, as evidenced by the peak at $\sim T_c/10$, 
the increase of $\tau_{\rm el{\mathchar"712D}ph}$ 
just below $T_c$ is expected to be small. 
This scenario is consistent with the slight change of 
$\kappa/T$ just below $T_c$. 

Below $T_p$, the scattering mechanism changes dramatically. 
In the lower inset of Fig.~2(a), we show $\rho(T)$ in two crystals at
high fields where $T_c(H)< T_p$. 
The normal-state $\rho$ well below $T_p$ exhibits the Fermi-liquid 
behavior $\rho_0+AT^2$ with $\rho_0=1.0~\mu\Omega$~cm and 
$A=0.14$-$0.20~\mu\Omega$~cm/K$^2$. 
This, together with the large $\gamma \sim 100$~mJ/K$^2$mol 
\cite{hiroi2,bruhwiler3}, 
follows the Kadowaki-Woods relation $A=a_{\rm KW}\gamma^2$ with $a_{\rm KW}$ 
close to the universal value of $10^{-5}~\mu \Omega$~cm(K~mol/mJ)$^2$ 
as shown in Fig.~3 \cite{tsuji}. 
This indicates that strongly correlated electrons 
with large mass are responsible for the $T^2$-dependence. 
Therefore it is natural to consider that below $T_p$ electron-phonon 
scattering suddenly diminishes and QP transport is dominated 
by electron-electron scattering, 
as schematically shown in the upper inset of Fig.~2(a);  
the enhancement of $\kappa/T$ stems from the enhancement 
of $\tau_{\rm el{\mathchar"712D}el}$. 
These results are also important for understanding the nature of 
the transition at $T_p$, namely our
results support the proposed rattling transition 
because the freezing of the K ion motion 
should strongly influence the phonon spectrum. 

To our knowledge, such enhancement of $\kappa_e$
in the superconducting state has been reported only in very clean 
high-$T_c$ cuprates \cite{yu} and 
heavy-fermion CeCoIn$_5$ \cite{kasahara}. 
We also note that the magnitude of the enhancement of $\tau_e$ observed 
in KOs$_2$O$_6$ is comparable to that in these systems, 
in which 
anisotropic $d$-wave pairing states are realized 
with strong electron correlations \cite{tsuei,izawa}.   

\begin{figure}[t]
\includegraphics[width=90mm]{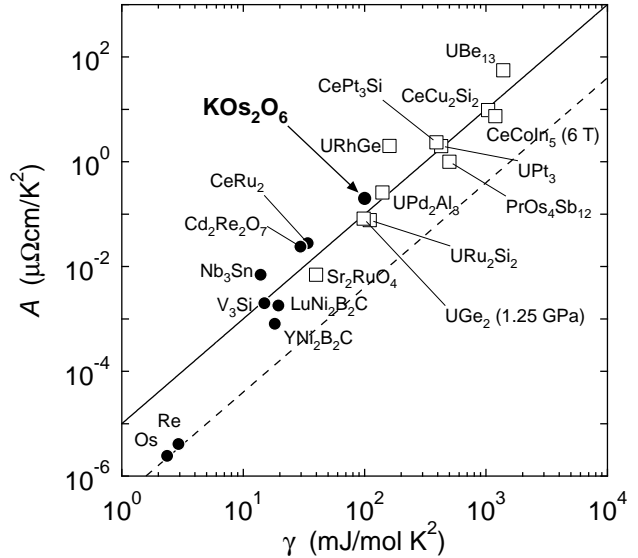}
\caption{
Coefficient $A$ vs Sommerfeld constant $\gamma$ 
(Kadowaki-Woods plot \cite{tsuji}) 
for various superconductors. Solid circles are for superconductors 
believed to have isotropic gap (or anisotropic $s$-wave symmetry). 
Open squares for anisotropic gap with nodes. 
The lines represent $A=a_{\rm KW}\gamma^2$ with $a_{\rm KW}=10^{-5}$ 
(solid) and $4\times 10^{-7}~\mu \Omega$~cm(K~mol/mJ)$^2$ (dashed). }
\end{figure}

Having established the evidence for strong electronic correlations, 
the next important issue is the paring symmetry. 
Owing to recent progress in understanding the thermal conductivity, 
it is well established that there is a fundamental difference 
in the heat transport 
between isotropic and anisotropic superconductors. 
We first discuss the gap function in terms of zero-field thermal 
conductivity at very low temperatures displayed in Fig.~2(b). 
As discussed above, we observe $\kappa/T=a+bT^2$ and 
the electronic contribution can be evaluated from 
the residual linear term   
$a=\kappa/T(T\rightarrow~0)\simeq 7\times 10^{-3}$ W/K$^2$m. 
In unconventional superconductors with a line node, 
the universal residual linear term $\kappa_{00}/T$ is present, 
which is independent of impurity concentration \cite{taillefer}. 
The question arises whether the finite $a$-term indicates a line node. 
To check this, we compare $a$ with the linear term in the normal state 
$\kappa_n/T$.  The Wiedemann-Franz law and $\rho_0=1.0~\mu\Omega$cm 
yield $\kappa_n/T=~$2.44~W/K$^2$m, a value larger than $a$ 
by a factor of more than 300. 
The residual linear term expected for a line node is given as 
$\frac{\kappa_{00}}{T}=\left( \frac{2}{\pi} \frac{\hbar}
{\Delta\tau_{\rm imp}} \right) \frac{\kappa_n}{T}$,
where $\Delta$ is the superconducting gap.
By using $\tau_{\rm imp}=\mu_0\lambda^2/\rho_0$ with 
$\lambda=270$~nm, and $\Delta = 1.9 k_{\rm B}T_c$ \cite{bruhwiler3},
$\kappa_{00}/T$ is estimated to be $ \sim$~0.07~W/K$^2$m. 
This value is an order of magnitude larger than $a$, 
indicating that low-temperature $\kappa$ in zero field 
is totally incompatible with 
the line node. The origin of the small residual linear term 
may be due to a non-superconducting 
metallic regime in the crystal.

\begin{figure}[tb]
\includegraphics[width=80mm]{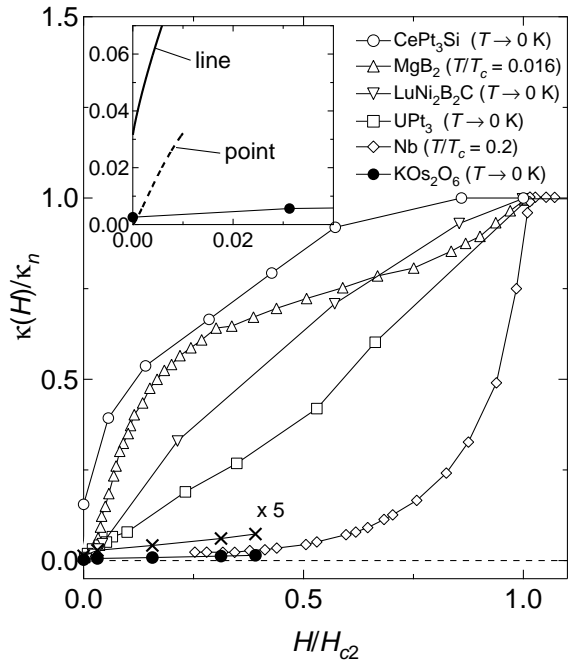}
\caption{ 
$H$-dependence of  $\kappa/\kappa_n$, as a function of $H/H_{c2}$. 
For comparison, data for the $s$-wave Nb, 
MgB$_2$ with two gaps, anisotropic CePt$_3$Si (line node), 
LuNi$_2$B$_2$C (anisotropic $s$), and UPt$_3$ are shown. 
Crosses indicate values for $\rho_0=5~\mu\Omega$cm. 
Inset: $H$-dependence of $\kappa$ at low fields. 
The solid and dashed lines represent the data for line 
and point nodes, respectively. }
\end{figure}

Strong evidence for fully gapped superconductivity is provided 
by the $H$-dependence of $\kappa$. 
Figure 4 depicts the $H$-dependence of $\kappa/\kappa_n$ 
in the low-temperature limit. 
For comparison, $\kappa(H)/\kappa_n$ for several superconductors, 
$s$-wave Nb, MgB$_2$ with two gaps \cite{mgb}, 
and anisotropic UPt$_3$, LuNi$_2$B$_2$C \cite{boaknin1}, 
and CePt$_3$Si (line node) \cite{izawa2}, are also plotted. 
We immediately notice that the $H$-dependence of $\kappa(H)/\kappa_n$ 
in KOs$_2$O$_6$ stays nearly constant up to $\sim0.4H_{c2}$ 
and is very close to that of Nb, in dramatic contrast with 
those in anisotropic superconductors. 
In the superconducting state, the thermal transport is governed 
by the delocalized QPs, which extend over the whole crystal. 
In $s$-wave superconductors the only QP states present at $T \ll T_c$ 
are those associated with vortices. 
When the vortices are far apart, these states are bound to 
the vortex core and are therefore localized and unable to transport heat; 
the conduction shows an exponential behavior with very slow growth 
with $H$ at $H \ll H_{c2}$, as observed in Nb. 
In sharp contrast, the heat transport in nodal superconductors 
is dominated by contributions from delocalized QP states. 
The most remarkable effect on the thermal transport is the Doppler shift 
of the energy of QPs \cite{kubert,point}. 
In the presence of line nodes where the DOS has 
a linear energy dependence ($N(E)\propto E$), $N(H)$ increases 
in proportion to $\sqrt{H}$. 
Therefore, the thermal conductivity in superconductors with 
large anisotropy grows rapidly at low field, 
as shown in UPt$_3$, LuNi$_2$B$_2$C, and CePt$_3$Si. 
In the inset of Fig.~4 is plotted $\kappa(H)/\kappa_n$ at low fields 
for gap functions with line \cite{kubert} and linear point \cite{point} 
nodes for unitary limit calculated using the same parameters 
as adopted in the analysis of the residual $\kappa/T$. 
It is clear that $\kappa(H)/\kappa_n$ of KOs$_2$O$_6$ shows much slower 
growth than that of superconductors with point and line nodes. 
Thus no discernible delocalized QPs exist at least for 
$H<0.4H_{c2}$ in KOs$_2$O$_6$. 
In case we have underestimated $\rho_0$ in the fit in the inset of Fig.~2, 
we also plot the case for $\rho_0$ 5 times larger and find that 
the slope of $\kappa/\kappa_n$ is still close to that of Nb. 
The marked insensitivity of $\kappa$ to $H$, together with the 
tiny residual linear term, leads us to conclude that the gap function 
of KOs$_2$O$_6$ is fairly isotropic. 

Strong electron correlations usually favor anisotropic 
superconductivity, since Coulomb repulsion prefers small 
probabilities in small pair distances. 
This is demonstrated in Fig.~3 by separating very anisotropic superconductors 
with nodes ($p$-, $d$-wave, etc.) from $s$-wave (and anisotropic $s$) 
superconductors by large $\gamma \gtrsim$ 30-40~mJ/mol K$^2$. Our
results therefore highlight the unusual full-gap superconductivity with 
strong correlations characterized by large $\gamma$ in KOs$_2$O$_6$. 

In summary, we have investigated the superconducting state 
of KOs$_2$O$_6$ by thermal conductivity. 
In contrast to previous studies, the isotropic gap function is 
firmly indicated. 
We also found the striking enhancement of the QP scattering time 
in the superconducting state. 
The present results indicate that strong electron correlations coexist with 
a fully gapped superconductivity, which
make KOs$_2$O$_6$ a quite unique system. 
How the unique superconductivity is related to geometrical frustration 
and rattling is an intriguing issue, which deserves further 
experimental and theoretical studies. 

We thank  K.~Behnia, S.~Fujimoto, H.~Ikeda, H.~Kontani, K.~Maki, and M.~Takigawa 
for valuable discussions.  This work was partly supported by Grants-in-Aid for Scientific Reserch from MEXT.  Y.K. was supported by the Research Fellowships of the Japan Society for the Promotion of Science for Young Scientists.

\end{document}